\documentclass[12pt]{article}
\usepackage{amsmath}
\usepackage{amsfonts}
\usepackage{amssymb}
\usepackage{latexsym}
\usepackage{graphicx}
\usepackage[english]{babel}
\input epsf.sty

\usepackage{indentfirst}
\usepackage{graphicx}

\begin{document}
\title{{\normalsize \hskip4.2in {USTC-ICTS-0726}} \\
{Some Issues Concerning Holographic Dark Energy}}
 \vspace{3mm}

 \author{{Miao Li$^{1,2}$, Chunshan Lin$^{1}$, Yi Wang$^{2,1}$}\\
 {\small $^{1}$  The Interdisciplinary Center for Theoretical Study, University of }\\
{\small Science and Technology of China (USTC), Hefei, Anhui 230027,
P.R.China} \\
{\small $^{2}$ Institute of Theoretical Physics, Academia Sinica,
Beijing 100080, P.R.China}}
\date{}
\maketitle

\begin{abstract}
We study perturbation of holographic dark energy and find it be
stable. We study the fate of the universe when interacting
holographic dark energy is present, and discuss a simple
phenomenological classification of the interacting holographic dark
energy models. We also discuss the cosmic coincidence problem in the
context of holographic dark energy. We find that the coincidence
problem can not be completely solved by adding an interacting term.
Inflation may provide a better solution of the coincidence problem.
\end{abstract}

\newpage

\section{Introduction}
 The cosmological constant problem\cite{ccref} is a longstanding problem in
theoretical physics, and has been taken more and more seriously
since the discovery of accelerated expansion of our
universe\cite{supernova}\cite{WMAP}\cite{SDSS}. In addition to the
problem why the cosmological constant is nonvanishing, there is also
the ``cosmic coincidence problem''\cite{ccref}.

Based on the validity of effective quantum field theory, Cohen
{\it et al} \cite{Cohen:1998zx} pointed out that the quantum
zero-point energy of a system should not exceed the mass of a black
hole of the same size. This observation relates the UV cutoff of a
system to its IR cutoff. As a cosmological application, Li
\cite{mli04} suggested to choose the future event horizon as the IR
cut-off, the energy density of vacuum is given by
\begin{equation}\label{rhod}
\rho_D=3c^2M_p^2R_h^{-2}~,
\end{equation}
where $R_h\equiv a \int_{t}^{\infty}dt'/a(t')$ is the size of future
event horizon. This is called holographic dark energy (HDE) and has
been studied extensively \cite{refHDE}.

As a phenomenological model, the stability of holographic dark
energy is an important issue and has been investigated by Myung
\cite{Myung:2007pn} first. Myung \cite{Myung:2007pn} assumed that
holographic dark energy is a usual fluid component. He
calculated the square of sound speed of holographic dark energy and
found it be negative, this leads to an instability of perturbation of
holographic dark energy. However, holographic dark energy is given
by the holographic vacuum energy, whose perturbation should be
treated globally. A calculation of perturbation of holographic dark
energy will be presented in sect.2. It is shown that perturbation of
holographic dark energy is stable, and we do not need to face the
negative sound speed square problem.

The interacting holographic dark energy, assuming that dark energy
interacts with matter, has become a popular topic recently
\cite{interactingref}. With the interacting term, the story of
holographic dark energy becomes more interesting. There is often an
attractor solution to the evolution equation, in which the effective
equations of state of dark energy and matter become identical in the
far future. In sect.3, we will give a simple and phenomenological
classification of the interacting terms, we will show that we can
tune the interacting parameter to avoid the phantom-like universe.

In sect.4, we shall discuss the coincidence problem of holographic
dark energy in a more natural way. We end this paper with conclusion
and discussion in sect.5.

\section{Stability of Holographic Dark Energy}

In this section, we investigate perturbation of holographic dark
energy. First, we shall calculate  perturbation of the future event
horizon, consequently we get the density perturbation of holographic
dark energy. Finally, we couple this density perturbation to
gravity. As an application, we analyze the coupled equation
approximately in the dark energy dominated era, and show that the
perturbation is stable. We also solve the perturbation equation
numerically outside the horizon. When dark energy dominates, the
solution agrees with the analytic result. When matter
dominates, the numerical result also shows that the perturbation is
stable.

\begin{figure}
\center
\includegraphics[width=10cm, height=7.5cm]{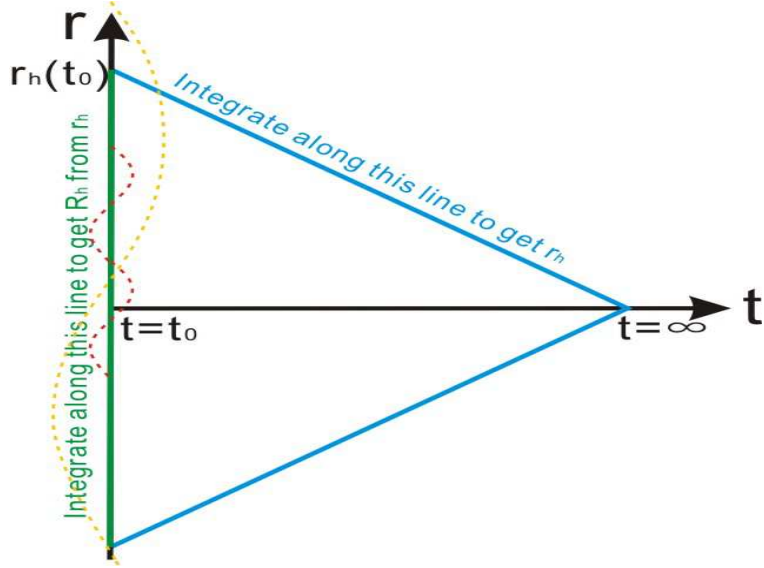}
\caption{\small{This figure illustrates how to calculate the
perturbation of the future event horizon. We first integrate along
the blue line, which is the geodesic for photons, to get the
coordinate distance of the future event horizon $r_h$, and then
integrate along the green line from $r=0$ to $r=r_h$ to get the
physical distance $R_h$. The dashed red and yellow curves
illustrates the sub-horizon and super-horizon perturbations,
respectively.}} \label{fig:hdepert}
\end{figure}

We consider the scalar type perturbation of the metric. In the
Newtonian gauge, the perturbated metric takes the form
\begin{equation}\label{metric}
  ds^2=-(1+2\Phi(r,t)) dt^2 + a(t)^2(1-2\Phi(r,t)) d{\bf x}^2~,
\end{equation}
where for simplicity, we have assumed that the perturbation is
spherically symmetric, $\Phi=\Phi(r,t)$, where $r=|{\bf x}|$. In
this metric, light traveling from the horizon towards the origin
$r=0$ still goes straightly. As illustrated in Fig.
\ref{fig:hdepert}, the future event horizon $R_h$ can be written as
\begin{equation}
  R_h(0,t)=\int_0^{r_h(t)} a(t) (1-\Phi(r,t))dr~,
\end{equation}
where $R_h(0,t)$ denotes the future event horizon at position $r=0$
at time $t$, $r_h(t)$ denotes the coordinate distance of the
future event horizon. At the first order in the perturbation theory,
$r_h\equiv r_{h0}+\delta r_h$, where $r_{h0}(t)=\int_t^\infty
dt'/a(t')$, and $\delta r_h$ can be written as
\begin{equation}
  \delta r_h(t)=\int_{t}^{\infty} \frac{2\Phi(r_{h0}(t'),t')}{a(t')}dt'~.
\end{equation}
So the variation of the future event horizon $R_h$ at the position
$r=0$ takes the form
\begin{equation}
  \delta R_h(0,t)\equiv R_h(0,t)-R_{h0}=a(t)\left\{
\int_t^\infty \frac{2\Phi(r_{h0}(t'),t')}{a(t')}dt'
-\int_0^{r_{h0}}\Phi(r,t)dr
  \right\}~.
\end{equation}
Note that for the background value, we have $R_{h0}= a r_{h0}$.

Using the definition of holographic dark energy (\ref{rhod}), and
varying $R_h$, we get the variation of the energy density of
holographic dark energy with respect to the metric perturbation
\begin{equation}
  \delta \rho_D=-2\rho_D \frac{\delta R_h}{R_h}~.
\end{equation}

\begin{figure}
\includegraphics[width=8cm]{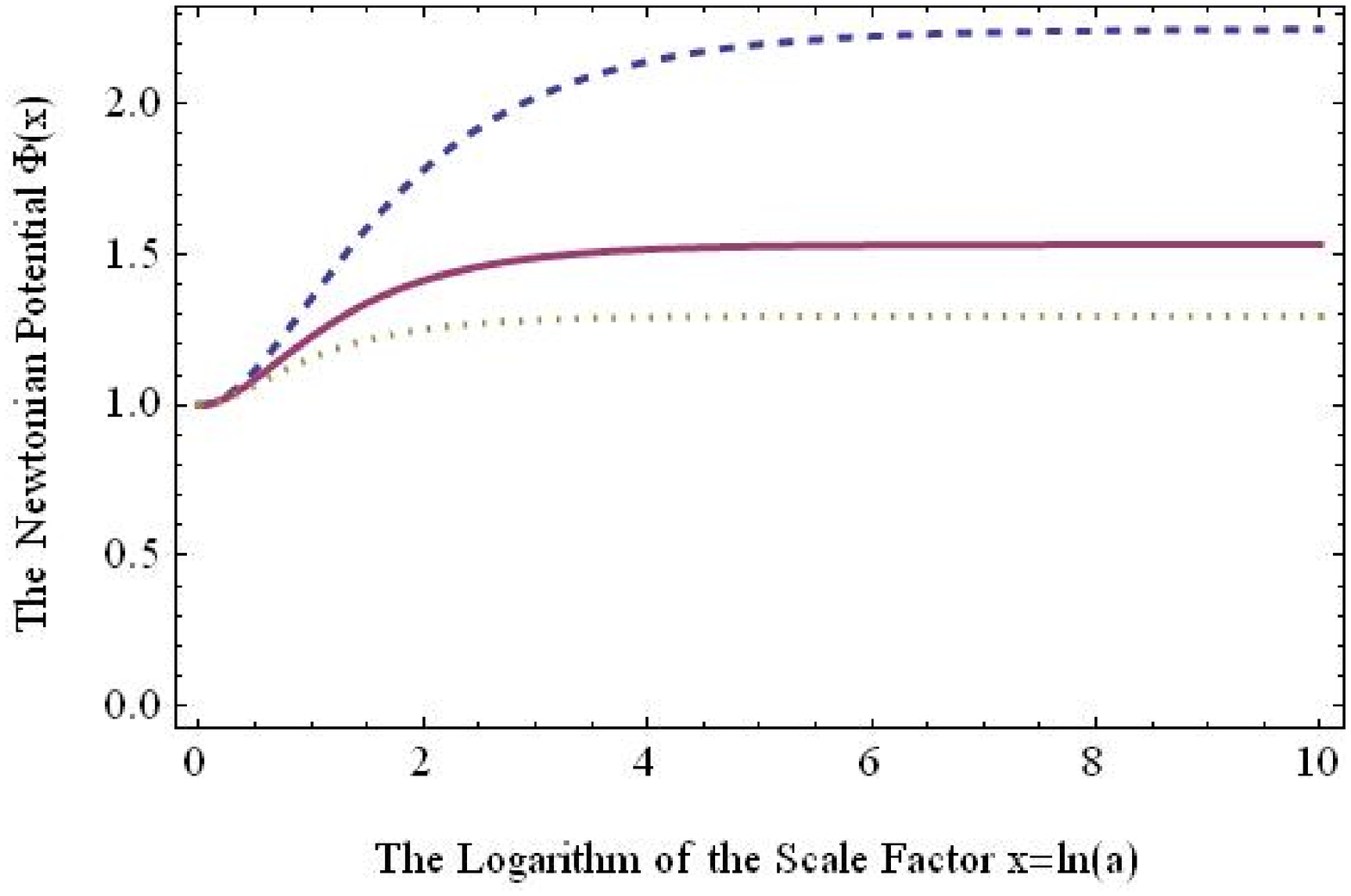}
\includegraphics[width=8cm]{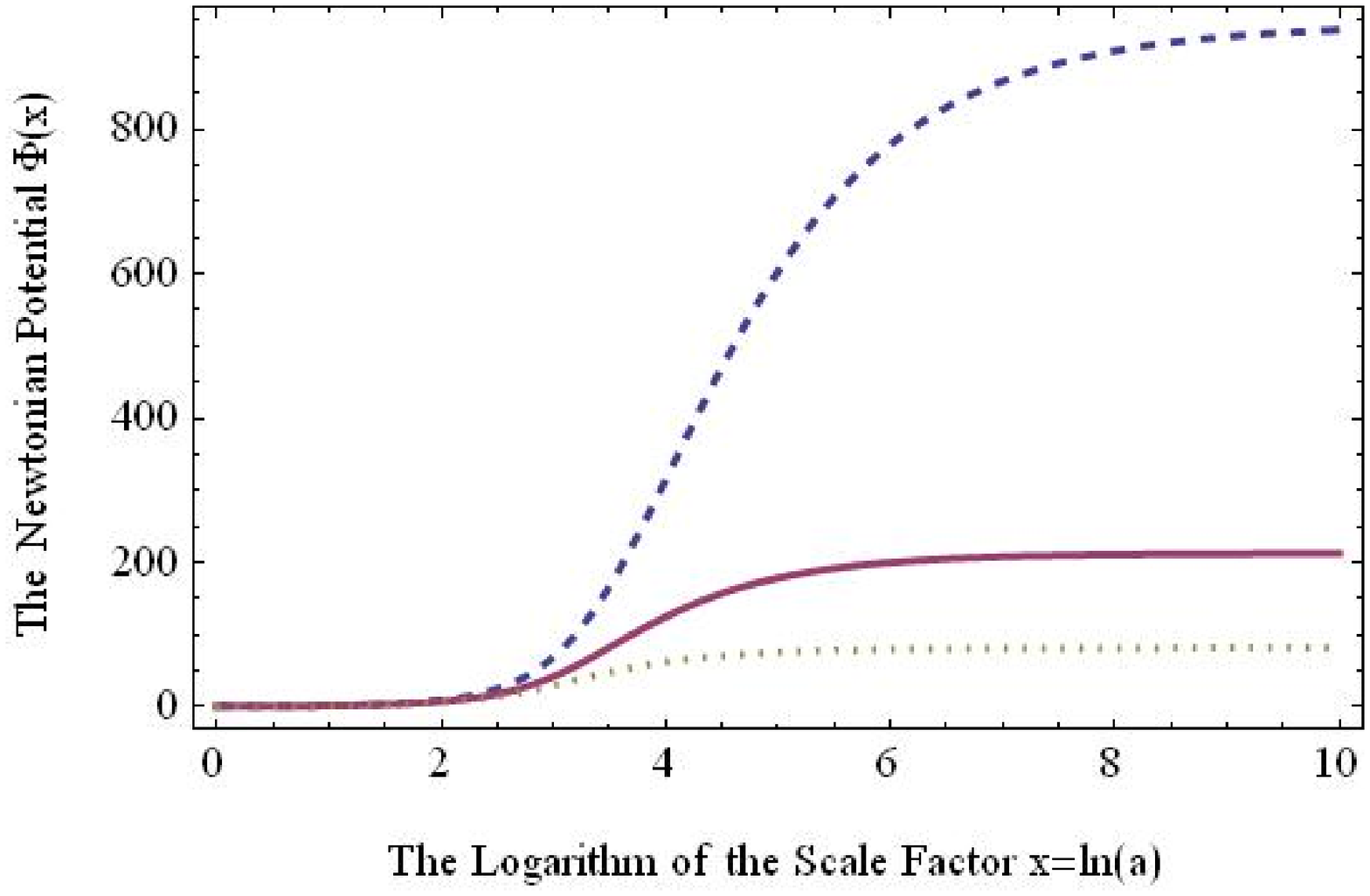}
\caption{\small{We plot the gravitational potential $\Phi$ as a
function of $x\equiv \ln a$ in the $k r_{h0}\ll 1$ case. The left
figure is plotted starting from the dark energy dominated era
$\Omega_D|_{x=0}=0.72$, and the right figure is plotted starting
from the matter dominated era $\Omega_D|_{x=0}=0.01$. The blue
(dashed), red (solid) and yellow (dotted) curves corresponds to
$c=0.8$, $c=1.0$ and $c=1.2$ respectively. In both figures, we see
that the perturbation approaches a constant mode. Note that we have
chosen the initial condition $\Phi(0)=0$ and $\Phi'(0)=1$. So
although in the right figure, the amplitude can grow ${\cal
O}(100)$} times before approaching the constant mode, but as the
initial condition should be set to $\Phi(0)\propto \rho_D$, the
physical amplitude is still not too large.} \label{fig:hdepertnum}
\end{figure}

Inserting this equation into the $00$-component of the perturbated
Einstein equation, one obtains
\begin{equation}
  \frac{\nabla^2}{a^2}\Phi-3H\dot\Phi-3H^2\Phi=\frac{1}{2M_p^2}(\delta\rho_D+\delta\rho_m)~.
\end{equation}
For simplicity, we neglect the matter density perturbation, so
$\delta\rho_m=0$. To solve this equation, we expand $\Phi$ using its
eigenfunction. Write
\begin{equation}
  \Phi(r,t)=\sum_k \Phi_k(t)\frac{\sin (kr)}{r}~,
\end{equation}
where we have dropped the $\cos(kr)/r$ terms, which lead to a
singularity at $r=0$. Then $\Phi_k(t)$ satisfies
\begin{eqnarray}\label{firstorder}
&&  \frac{M_p^2}{\rho_D} \frac{\sin
(kr)}{r}r_{h0}(t)\left\{\frac{k^2}{a^2}\Phi_k(t)+ 3H \dot\Phi_k(t) +
3H^2
  \Phi_k(t)\right\}\nonumber\\
&=&
  \int_t^\infty \frac{2\Phi_k(t')\sin(kr_{h0}(t'))dt'}{a(t')r_{h0}(t')} -\Phi_k(t)\int_0^{r_{h0}(t)}\frac{\sin(kr)}{r}dr~.
\end{eqnarray}
One way to deal with this equation is to take derivative with
respect to $t$. This integral equation becomes a differential
equation
\begin{eqnarray}
  \ddot\Phi_k &+& \frac{1}{3H}\left\{\left(3\dot H+9H^2-\frac{9H}{R_{h0}}+\frac{k^2}{a^2}\right)+\frac{a \rho_D}{M_p^2
  R_{h0}}\int_0^{r_{h0}}\frac{\sin(kr)}{kr}dr\right\}\dot\Phi_k\nonumber\\
&+&\frac{1}{3H}\left\{\left(6H\dot H+6
H^3-\frac{3}{R_{h0}}\frac{k^2}{a^2}-\frac{9H^2}{R_{h0}}\right)+\frac{\rho_D}{M_p^2
R_{h0}} \frac{\sin(kr_{h0})}{kr_{h0}}\right\}\Phi_k=0~.
\end{eqnarray}
This equation can be solved at least numerically. As an example, the
evolution of the $kr_{h0}\ll 1$ mode is shown in Fig.
\ref{fig:hdepertnum}.

Eq.(\ref{firstorder}) can also be treated directly in the
dark energy dominated era $\rho_D>\rho_m$. To investigate the
stability of the perturbation mode, we focus on behavior of
$\dot\Phi_k/\Phi_k$. When $\dot\Phi_k/\Phi_k\rightarrow 0$, the
perturbation mode is frozen, and when $\dot\Phi_k/\Phi_k<0$, the
perturbation mode is decaying.

Let us first consider the super-horizon mode $kr_{h0}(t)\ll 1$. In
this case, Eq. (\ref{firstorder}) can be written as
\begin{equation}
  3H\dot\Phi_k\simeq -3H^2\Phi_k-\frac{\rho_D}{M_p^2}\Phi_k +\frac{2\rho_D}{M_p^2
  r_{h0}}\int_{t}^{\infty}\frac{dt'}{a(t')}\Phi_k(t')~.
\end{equation}
The integral in the last term can be estimated using
\begin{equation}
  \int_{t}^{\infty}\frac{dt'}{a(t')}\Phi_k(t')\simeq
  \int_{t}^{\infty}\frac{dt'}{a(t')}\left(\Phi_k(t)+\dot\Phi_k(t) (t'-t)\right)~.
\end{equation}
Note that for $\int_t^\infty \frac{dt'}{a(t')} (t'-t)$, using
$dt'/a(t')=-dr_{h0}$ and expanding $R_{h0}(t')\simeq
R_{h0}(t)+(t'-t)\dot R_{h0}(t)$, we have
\begin{equation}
  \int_t^\infty \frac{dt'}{a(t')} (t'-t)
  \simeq r_{h0}R_{h0}+\int_{t}^\infty \frac{dt'}{a(t')}
  (t'-t)\left(c\sqrt{\frac{\rho_c}{\rho_D}}-1\right)~,
\end{equation}
where $\rho_c$ is the critical density. So we see that for the dark
energy dominated era and $c\approx 1$, the expansion works well. Up
to the leading order, $\int_t^\infty \frac{dt'}{a(t')} (t'-t)\simeq
r_{h0}R_{h0}$. So Eq. (\ref{firstorder}) can be written as
\begin{equation}
  (2c\sqrt{\frac{\rho_D}{\rho_c}}-1)\dot\Phi_k=H(1-\frac{\rho_D}{\rho_c})\Phi_k~.
\end{equation}
During the dark energy domination, $1-\frac{\rho_D}{\rho_c}$
approaches zero quickly, so the super-horizon perturbation
approaches a constant. There is no instability for the
perturbation.

  Note that when $c \simeq 1/2$, there exists a parameter region where
  $2c\sqrt{\rho_D/\rho_c}-1$ approaches zero before $\rho_D\rightarrow
  \rho_c$. In this case, the next to leading order correction to
  (13) should be considered. However, the experimental data
  indicates that $c$ does not lie in this regime.

For the sub-horizon mode $kr_{h0}(t)\gg 1$, similar analysis can be
performed. One can divide the nonlocal integration in
(\ref{firstorder}) into two parts, namely $kr_{h0}(t')\gg 1$ and
$kr_{h0}(t')\ll 1$, and use $|\sin[kr_{h0}(t')]|<1$ and
$|\sin[kr_{h0}(t')]|<kr_{h0}(t')$ respectively. It can be shown that
the dominant contribution comes from
\begin{equation}
  3H\dot\Phi_k\simeq -\left(\frac{k^2}{a^2}+3H^2\right)\Phi_k~,
\end{equation}
and the other terms are suppressed by a factor of $1/[kr_{h0}(t)]$.
So the sub-horizon mode is a decaying mode. Again, no
instability appears.

Before proceeding to the next section, we would like to discuss two
physical issues of the holographic dark energy.

First, from the calculation above, we see clearly that perturbation
of holographic dark energy is nonlocal. This is completely different
from a usual fluid component. For a usual fluid component, the
perturbation equation has a non-vanishing local limit. In this
limit, the perturbation equation follows from local conservation of
the energy-momentum tensor $\partial_\mu T^{\mu\nu}=0$, and the
sound speed $c_s$ for the perturbation is given by $c_s^2=dp/d\rho$.
When $c_s^2=dp/d\rho<0$, the perturbation of the fluid is unstable.
But for holographic dark energy, the perturbation of the energy
density comes from the perturbation of the metric, and does not
suffer such instability.

Second, although we discussed the evolution of perturbation for
holographic dark energy, we did not discuss the initial condition
for it. As it is difficult to write the holographic dark energy
component into the Lagrangian, the quantum initial condition for
holographic dark energy is not available. Another source for
perturbation of holographic dark energy is perturbation of the
matter component. Perturbation of the matter component couples
to the metric perturbation, thus providing a initial condition for the
holographic dark energy perturbation.

\section{The Fate of Interacting Holographic Dark Energy}

In this section we will make a simple and phenomenological
classification of interacting holographic dark energy. We also study
the fate of the universe with interacting holographic dark energy:
in what case it will be phantom-like, in what case phantom will be
avoided, and whether the big rip will happen or not.

For simplicity, use $w$ to denote the effective index of the
equation of state of dark energy (which is sometimes written as
$w_D^{\rm eff}$ in the literature), and use $w_m$ to denote the
effective index of the equation of state of matter (which is
sometimes written as $w_m^{\rm eff}$ in the literature) in the
following discussion. Please note that we neglect the curvature
of the universe, the following calculation was done with the
assumption that the universe is flat.
\subsection{Dark Energy Decay to Matter}
First, we consider the case when holographic dark energy decays to
matter
\begin{equation}\label{evolution1}
  \rho_D'+3(1+w_D)\rho_D=3b\rho_D~,
\end{equation}
  \begin{equation}\label{evolution2}
  \rho_m'+3\rho_m=-3b\rho_D~,
\end{equation}
where the prime denotes derivative with respect to $\ln a$. It is
worthy to note that the total energy is conserved in the interacting
holographic dark energy model, although dark energy and matter are
not conserved separately. For the lack of the first principle of
holographic dark energy, we take the above equation
phenomenologically. We simply follow other works done about the
interacting holographic dark energy\cite{interactingref}. The decay
rate is proportional to the energy density of dark energy, so
naturally we have $b<0$, meaning that dark energy decays to matter.
Moreover, $b>0$ will lead to unphysical consequence in physics, such
as $\rho_m$ will become negative and $\Omega_D$ will be larger than
1 in the future. So we assume $b<0$ in this subsection.

Comparing with the effective equation of state
\begin{equation}
\rho_D'+3(1+w)\rho_D=0~,~~~\rho_m'+3(1+w_m)\rho_m=0~,
\end{equation}
we find the indices of the effective equation of state
\begin{equation}\label{wm1}
w=w_D-b ~,~~~w_m=b\Omega_D/\Omega_m~.
\end{equation}
If the index of the effective equation of state of dark energy
satisfies $w<-1$, dark energy is phantom-like.

 Taking
derivative of Eq.(\ref{rhod}) with respect to $\ln a$, we have
\begin{equation}\label{rho1}
\rho_D'=2\rho_D\left(\frac{\sqrt{\Omega_D}}{c}-1\right)~,
\end{equation}
from Eqs.(\ref{evolution1})(\ref{rho1}), we get
\begin{equation}
w_D=-\frac {1}{3}-\frac {2}{3}\frac {\sqrt{\Omega_D}}{c}+b~.
\end{equation}
Using the definition of $\Omega_D$ and taking derivative of
$\Omega_D$ with respect to $\ln a$, we have
\begin{equation}\label{omegad1}
\Omega_D'=-2\Omega_D+\frac{2\Omega_D^{3/2}}{c}-2\Omega_D
\frac{H'}{H}~.
\end{equation}
From Eq.(\ref{omegad1}), we get
\begin{equation}
\frac {H'}{H}=-\frac
{\Omega_D'}{2\Omega_D}-1+\frac{\sqrt{\Omega_D}}{c}~.
\end{equation}
From the Friedmann equation
\begin{equation}\label{Hubble1}
\dot{H}=-4\pi G(\rho +p)=-4\pi G(\rho +\rho_D w_D+\rho_r w_r)~,
\end{equation}
we get
\begin{equation}\label{hubble2}
\frac{H'}{H}=\frac{\Omega_D}{2}+\frac{\Omega_D^{3/2}}{c}-\frac{3}{2}b\Omega_D
-\frac{3}{2}~,
\end{equation}
the last term of the RHS of Eq.(\ref{Hubble1}) can be neglected.

Substituting Eq.(\ref{hubble2}) into Eq.(\ref{omegad1}), we obtain
the differential equation for $\Omega_D$
\begin{equation}\label{omegad'}
\frac{\Omega_D'}{\Omega_D}=(1-\Omega_D)(1+\frac{2\sqrt{\Omega_D}}{c})+3b\Omega_D~.
\end{equation}

\begin{figure}
\includegraphics[width=\textwidth]{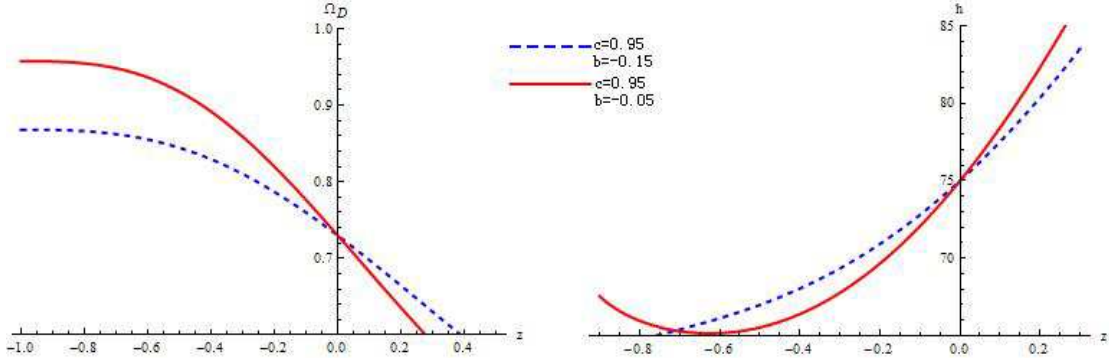}
\caption{\small{This figure illustrates evolution of the universe
with assumption that holographic dark energy can decay to matter.
The red (solid) curve corresponds to the phantom-like universe, and
blue (dashed) curve corresponds to the other case. We choose initial
condition $\Omega_{D0}=0.73$, $h_0=75$ in our numerical
calculation.}} \label{fig:detomat}
\end{figure}

We solve Eqs.(25)(26) numerically and the evolution of the universe
has been shown in Fig. \ref{fig:detomat}. Now we will discuss these
equations analytically.

Considering Eq.(\ref{omegad'}), we find that the LHS of
Eq.(\ref{omegad'}) will vanish only when the scale factor goes to
infinity. To see this, we define
\begin{equation}f(y)\equiv\frac{2y'}{y}=(1-y^2)(1+\frac{2y}{c})+3by^2~,
\end{equation}
where $y\equiv\sqrt{\Omega_D}$. The equation $f(y)=0$ has three
roots. Since $f(0)=1>0$ and $f(1)=3b<0$, there is one root in the
region [0,1] at least. We only consider the region $[0,1]$ since it
is the physical region, for Friedmann equation in the flat universe,
the energy density of holographic dark energy and that of matter should
be positive. So $\Omega_D$ should never go beyond the region [0,1].
We assume that $y_1$ is the first root in the physical region
$[0,1]$. Thus, we find the integral equation
\begin{equation}\label{integ}
\int_{\varepsilon}^{y_1}\frac{dy}{y\prod_{i=1}^{3}(y-y_i)}=\ln{a}-\ln{a(\varepsilon)}~,
\end{equation}
where $\varepsilon$ is a cut-off at the early universe. We
find that the LHS of Eq.(\ref{integ}) diverges, which means that the
scale factor approaches infinity as $y\rightarrow y_1$. Combining the above equation
with the fact that $\Omega_D'>0$ when $\Omega_D\rightarrow 0$, we
conclude that $\Omega_D'$ will remain positive, and approaches zero
when the scale factor approaches infinity.

When the index of the effective equation of state satisfies
$w=w_D-b\geq-1$, {\it i.e.} $\frac{\sqrt{\Omega_D}}{c}\leq1$, dark energy
does not behave like phantom. As we have discussed, $\Omega_D$
is an increasing function with the scale factor. So once the no
phantom condition $\frac{\sqrt{\Omega_D}}{c}\leq1$ is satisfied at
$a\rightarrow \infty$, it will be satisfied along the whole history
of the universe. By setting $\Omega_D'=0$ in Eq.(\ref{omegad'}), we
get $b\leq1-c^{-2}$. It is the necessary condition to avoid the
phantom phase.

We can also prove that $b\leq1-c^{-2}$ is the sufficient condition
of no phantom. To see this, we first investigate the limiting case
$b=1-c^{-2}$, then use the monotonicity of $\Omega_D$ in $b$ to prove for
the general case. Substituting $b=1-c^{-2}$
into $f(y)=0$, we get three roots, $y_1=c$,
$y_{2,3}=\frac{-3\pm \sqrt{9-8c^2}}{4c}$. Note that $y_{2,3}<0$, so
$y_1$ is the only root making sense physically. Note that
solution $y_1=c$ corresponds to $w=-1$, which is the boundary of the
no phantom condition.

To see the monotonicity, suppose $y_i$ is the root of $f(y)=0$,
\begin{equation}\label{yi1}
(1-y_i^2)(1+\frac{2y_i}{c})+3by_i^2=0~.
\end{equation}
Taking derivative of the above equation with respect to $b$, we get,
\begin{equation}\label{yb1}
\frac{dy_i}{db}=\frac{-3y_i^2}{-2y_i+\frac{2}{c}-\frac{6y_i^2}{c}+6by_i}~.
\end{equation}
There is only one inflexion, $\frac{dy_i}{db}=0$, $y_i=0$, so the
function varies monotonously when $y_i\geq0$. Substitute
$b=1-c^{-2}$ into Eq.(\ref{yb1}), one finds $\frac{dy_i}{db}>0,$
around $y_i \rightarrow c$. That means $y=c$ is largest root of
Eq.(\ref{yi1}). So we conclude that for general $b$ satisfying
$b\leq1-c^{-2}$, we have $\frac{\sqrt{\Omega_D}}{c}\leq1,$ and
phantom will be avoided. Thus, we conclude that $b\leq1-c^{-2}$ is
the sufficient and necessary condition of no phantom.

In dark energy models, phantom usually causes the big rip. In the
remainder of this subsection, we will verify that this statement is
also true for the interacting holographic dark energy. First, we
will prove that if the no phantom condition $b\leq1-c^{-2}$ is
satisfied, there will be no big rip.

From equation
\begin{equation}\label{dotH1}
\dot{H}=-4\pi G[(1+w)\rho_D+(1+w_m)\rho_m],
\end{equation}
where
$w_m=\frac{b\Omega_D}{\Omega_m}\geq-\frac{1}{3}(1+\frac{2\sqrt{\Omega_D}}{c})\geq-1$,
we get $\dot{H}\leq0$, the Hubble parameter will become smaller
and smaller and big rip will never happen.

On the other hand, if dark energy is phantom-like, we can show
that the big rip will definitely happen. Consider the asymptotic
behavior of the evolution equation. When $a\rightarrow\infty$, to
have phantom, we have $1+w=1+w_m=-\alpha$, where $\alpha$
is a positive constant. Eq.(\ref{dotH1}) can be rewritten as,
$\dot H=-4\pi G[-\alpha (\rho_D+\rho_m)]=\frac{3\alpha}{2}H^2$ in
the $a\rightarrow \infty$ limit, and $H=\frac{1}{H_0-3\alpha t/2}$,
where $H_0$ is a integral constant. So the big rip happens in a
finite time $t=\frac{3}{2}\frac{H_0}{\alpha}.$

\subsection{Hybrid Interaction}

For the hybrid interaction, the interacting term is proportional to
the critical energy density. The evolution equations are
\begin{equation}
\rho_D'+3(1+w_D)\rho_D=3b\rho_c~,
\end{equation}
\begin{equation}
\rho_m'+3\rho_m=-3b\rho_c~.
\end{equation}
Dark energy will become dominant in this case, and
$\rho_D\rightarrow\rho_c$. For the same reason as in Subsection 3.1,
we consider the case $b<0$.

\begin{figure}
\includegraphics[width=\textwidth]{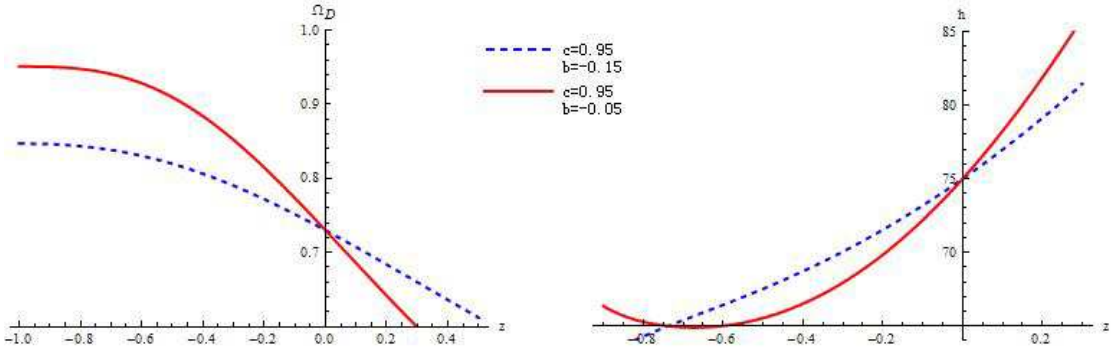}
\caption{\small{This figure illustrates evolution of the universe
with assumption that the interacting term is proportional to the
critical energy density. The red (solid) curve corresponds to the
phantom-like universe, and blue (dashed) curve corresponds to the
other case. We find that evolution is very similar with the case
dark energy decay to matter.}} \label{fig:hyint}
\end{figure}

With a same procedure as in the previous subsection, we obtain
differential equations for $\Omega_D$ and $H$
\begin{equation}\label{omegad'2}
\frac{\Omega_D'}{\Omega_D}=(1-\Omega_D)(1+\frac{2\sqrt{\Omega_D}}{c})+3b~,
\end{equation}
\begin{equation}
\frac{H'}{H}=\frac{\Omega_D}{2}+\frac{\Omega_D^{3/2}}{c}-\frac{3}{2}b-\frac{3}{2}~.
\end{equation}
Solving above two equations numerically, We get the evolution of the
universe, which has been shown in Fig. \ref{fig:hyint}. The indices
of the effective equations of state of holographic dark energy and
matter are
\begin{equation}\label{wm2}
w=-\frac{1}{3}-\frac{2\sqrt{\Omega_D}}{3c}~,~~~w_m=\frac{b}{\Omega_m}~.
\end{equation}
Using Eq.(\ref{omegad'2}) and the condition
$w=-\frac{1}{3}-\frac{2\sqrt{\Omega_D}}{3c}\geq -1$, we get the sufficient condition of no phantom is
$b\leq c^2-1$. This condition is also the necessary condition, as can be shown in
the same way as in the previous subsection. We rewrite Eq.(\ref{omegad'2})
in terms of $y\equiv\sqrt{\Omega_D},$
\begin{equation}
f(y)\equiv\frac{2y'}{y}=(1-y^2)(1+\frac{2y}{c})+3b~.
\end{equation}
 Assume that
$y_i$ is the root of $f(y)=0$,
\begin{equation}\label{yi2}
(1-y_i^2)(1+\frac{2y_i}{c})+3b=0~.
\end{equation}
There are three roots in the limit case $b=c^2-1$,$y_1=c$,
$y_{2,3}=\frac{-3c\pm\sqrt{16-15c^2}}{4}$.

In addition, Eq.(\ref{integ}) implies that there is a constraint on
$b$. From Eq.(\ref{integ}), we know that holographic dark energy
evolves monotonously along the whole history of the universe. Take into
account the fact that $\Omega_D$ is close to zero at the beginning of evolution
of the universe, $\Omega_D'$ must be positive all the time. To meet
this requirement, we need $b>-1/3$. Otherwise, $\Omega_D$ will be a
decreasing function in time. Combining this with Eq.(\ref{yi2}), we obtain
$\frac{2}{c}<\frac{2y_i}{c}+y_i$.

 Taking derivative of Eq.(\ref{yi2}) with respect to b, we get
\begin{equation}\label{yb2}
\frac{dy_i}{db}=\frac{-3}{-\frac{6y_i^2}{c}-2y_i+\frac{2}{c}}~.
\end{equation}
We see that the denominator is negative  since
$\frac{2}{c}<\frac{2y_i}{c}+y_i$. So $y_i$ increases monotonously
with respect to $b$ in the physical region. If we tune $b$ such that $b\leq
c^2-1$, we have
$w=-\frac{1}{3}-\frac{2}{3}\frac{\sqrt{\Omega_D}}{c}\leq-1$ and
phantom phase will be avoided.

By a similar analysis as in the previous subsection, we find that $b\leq c^2-1$ is
also the sufficient and necessary condition of no big rip.

\subsection{Matter Decay to Dark Energy}
Finally, we consider the case when matter decays to dark energy,
\begin{equation}
\rho_D'+3(1+w_D)\rho_D=3b\rho_m~,
\end{equation}
\begin{equation}
\rho_m'+3\rho_m=-3b\rho_m~.
\end{equation}
In this case, $b\geq0$, and the differential equation for $\Omega_D$
takes the form
\begin{equation}\label{omega'3}
\frac{\Omega_D'}{\Omega_D}=(1-\Omega_D)(1+\frac{2\sqrt{\Omega_D}}{c})+3b(1-\Omega_D)~.
\end{equation}
The differential equation for H can be written as
\begin{equation}
\frac{H'}{H}=\frac{\Omega_D}{2}+\frac{\Omega_D^{3/2}}{c}-\frac{3b}{2}(1-\Omega_D)-\frac{3}{2}~.
\end{equation}
\begin{figure}
\includegraphics[width=\textwidth]{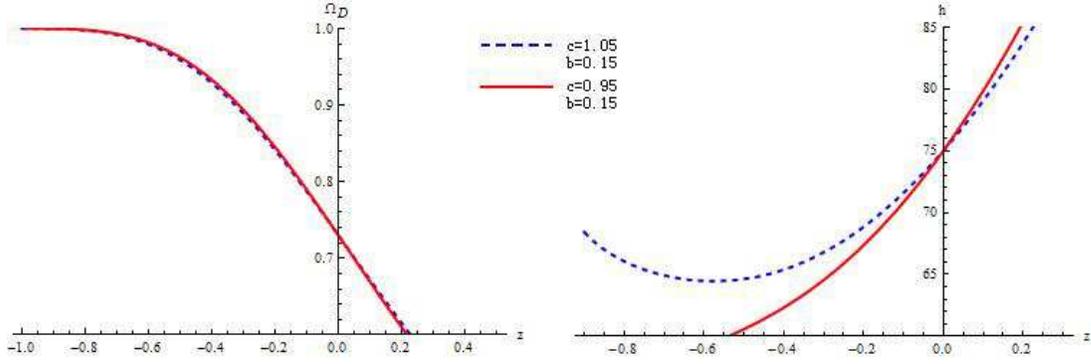}
\caption{\small{This figure illustrates evolution of the universe with
assumption that matter can decay to dark energy. The red (solid)
curve corresponds to the phantom-like universe, and blue (dashed)
curve corresponds to the other case. We find that dark energy will be
dominant and $\Omega_D=1$ eventually.}} \label{fig:matode}
\end{figure}
We obtain the indices of the effective equation of state
\begin{equation}\label{wm3}
w=-\frac{1}{3}-\frac{2\sqrt{\Omega_D}}{3c}~,~~~w_m=b~.
\end{equation}

Taking $\Omega_D'$ in Eq.(\ref{omega'3}), there is only
one root in the range [0,1], dark energy will be dominant, and
$\Omega_D$ approaches $1$ eventually. The sufficient and necessary condition of
no phantom is $c\geq 1$. It is also the sufficient and necessary
condition of no big rip. The analysis is just a little different from the
previous ones, because $w_m=b$ all the time and there is no attractor
solution to the evolution equation of state. Consider equation,
\begin{equation}\label{dotH2}
\dot{H}=-4\pi G[(1+w)\rho_D+(1+b)\rho_m],
\end{equation}
If no phantom, {\it i.e.} $1+w\geq0$, $\dot{H}<0$ all the time, and
the big rip will never happen. If dark energy is phantom-like,
$c<1$, Note $\Omega_D\rightarrow1$ eventually, we can neglect the
second term in Eq.(\ref{dotH2}), $H=\frac{1}{H_0-(c^{-1}-1)t}$ when
$t=\frac{H_0}{c^{-1}-1}$, the big rip will occur. The evolution of
the universe in this case has been shown in Fig. \ref{fig:matode}.

\section{The Coincidence Problem}

The usual solution to the coincidence problem is to calculate the
ratio of duration of coincidence state and lifetime of the universe,
assuming that dark energy is phantom-like and the universe will end with the
big rip \cite{coincidence}. The coincidence problem is solvable if
this ratio is not too small. Another solution to  the coincidence
problem is the interacting dark energy
\cite{interactingref}. We shall show that the interacting dark energy does not solve
the coincidence problem in this section.
We consider the solution
proposed by Li in the original paper of holographic dark energy
\cite{mli04} a more natural solution.

We start with evolution equations
\begin{equation}\label{27}
\rho_D'+3(1+w)\rho_D=0~,
\end{equation}
\begin{equation}\label{28}
\rho_m'+3(1+w_m)\rho_m=0~,
\end{equation}
where the prime denotes derivative with respect to $\ln a$, and $w$,
$w_m$ are the effective indices of the equations of state of holographic dark energy
and matter, respectively.

We rewrite Eq.(\ref{27}) in terms of an integral form
\begin{equation}
\ln \rho_D/ \rho_{D0}=\int_0^{\ln a}-3(1+w)d\ln a'
\end{equation}
Recall the median law of integral, we write the median value of $w$
as a constant $\tilde{w}$ varying in the interval ($-\frac{1}{3}$,
$-\frac{1}{3}-\frac{2}{3c}$). So the above integration can be
written as $\ln \rho_D/\rho_{D0}=-3(1+\tilde{w})\ln a$, we get
$\rho_D=\rho_{D0}a^{-3(1+\tilde{w})}$. Similarly we write the median
value of $w_m$ as a constant $\tilde{w}_m$, and we get
$\rho_m=\rho_{m0}a^{-3(1+\tilde{w}_m)}$, where $\rho_{D0}$ and
$\rho_{m0}$ are energy densities at the time where we set the scale
factor $a_0=1$.

From the Friedmann equation
\begin{equation}
H^2=\frac{8\pi G}{3}(\rho_m+\rho_D)~,
\end{equation}
and the definition of the Hubble parameter $H=\frac{d\ln a}{dt}$,
the age of the universe can be written in an integral form as
follows
\begin{equation}
t=\int_0^{a_0}[\frac{8\pi G}{3}(\rho_m+\rho_D)]^{-1/2}d\ln a~.
\end{equation}
Rewrite it in terms of $r\equiv\frac{\rho_D}{\rho_m}$, we obtain
\begin{equation}\label{39}
t=H_0^{-1}(1+r_0)^{1/2}r_0^{\frac{1+\tilde{w}}{2(\tilde{w}-\tilde{w}_m)}}\frac{1}{-3(\tilde{w}-\tilde{w}_m)}\int_0^{r_0}(1+r)^{-1/2}
r^{-\frac{1+\tilde{w}_m}{2(\tilde{w}-\tilde{w}_m)}-1}dr~,
\end{equation}
The integration in Eq.(\ref{39}) satisfies the follow inequality
\begin{equation}
\int_0^{r_0}(1+r_0)^{-1/2}r^{-\frac{1+\tilde{w}_m}{2(\tilde{w}-\tilde{w}_m)}-1}dr<\int_0^{r_0}(1+r)^{-1/2}
r^{-\frac{1+\tilde{w}_m}{2(\tilde{w}-\tilde{w}_m)}-1}dr~,
\end{equation}
and
\begin{equation}
\int_0^{r_0}(1+r)^{-1/2}
r^{-\frac{1+\tilde{w}_m}{2(\tilde{w}-\tilde{w}_m)}-1}dr<\int_0^{r_0}r^{-\frac{1+\tilde{w}_m}{2(\tilde{w}-\tilde{w}_m)}-1}~.
\end{equation}
So we get
\begin{equation}
\frac{2}{3}H_0^{-1}(1+\tilde{w}_m)^{-1}<t<\frac{2}{3}H_0^{-1}(1+r_0)^{1/2}(1+\tilde{w}_m)^{-1}~.
\end{equation}
If we input the value of $r_0$ at the present
time \cite{WMAP}\cite{SDSS}, we find that $t$ is about
$H_0^{-1}(1+\tilde{w}_m)^{-1}$. From the previous section, we know that
$w_m$ is proportional to $b$, the interacting parameter. We
see the coefficient of $H_0^{-1}$ is $b$ dependent. For $b=0$, $t$
is about $H_0^{-1}$. For $b\neq0$, from Eq.(\ref{wm1}),
$w_m=\frac{b\Omega_D}{\Omega_m}$  for
dark energy decaying to matter case, and  from Eq.(\ref{wm2}), $w_m=\frac{b}{\Omega_m}$
for hybrid interaction case,
where $b$ is always negative. From Eq.(\ref{wm3}), we have
$w_m=b$ $(b>0)$ for the case in which matter decaying to dark energy.

The key issue of the coincidence problem is why the ratio of
holographic dark energy to matter is order one nowadays, in other
words, why the ratio is order one when the age of universe is about
$10^{10}$ years. The order of magnitude of the age of the universe
is determined by $H_0^{-1}$. The relationship between the age of the
universe and $H_0^{-1}$ depends on the initial condition of the
universe. The interacting term can change the coefficient in the front of
$H_0^{-1}$, but it can not provide any information about the value of
$H_0^{-1}$. Thus, the interacting holographic dark energy can not solve the
coincidence problem completely in this sense. The above approach
just solves the coincidence problem partially.

To solve the coincidence problem completely, the initial density of
holographic dark energy and the influence of inflation should be
taken into account. As proposed in the original paper of
holographic dark energy \cite{mli04}, the initial energy density of
holographic dark energy has been inflated away by a factor exp$(-2N)$ in the
inflation epoch, where $N$ is the e-folding number of inflation. So
the ratio between $\rho_D$ and $\rho_r$, the radiation density,
should be about $10^{-52}$ at the onset of the radiation dominated
epoch, if we suppose that the inflation energy scale be $10^{14}$
$GeV$ and inflaton energy completely decays into radiation at the
end of inflation. This will lead to the order 1 ratio of holographic dark
energy to matter in our epoch. Thus inflation not only solves the
traditional naturalness problems and helps to generate primordial
perturbations, but also solves the cosmic coincidence problem.

\section{Conclusion and Discussion}
In this paper, we study perturbation of holographic dark energy.
Since holographic dark energy is just the
holographic vacuum energy, its perturbation is global. We
calculate perturbation of the holographic dark energy, and find
it stable.

Many numerical and  analytic works have been done on the interacting
holographic dark energy. We made a simple and phenomenological
classification of interacting holographic dark energy in this paper,
and derived the sufficient and necessary condition of no phantom
(the big rip). Needless to say, this classification has not been
done previously. It is worth to note that we write the interacting
term just by hand for lack of knowledge of the first principle of
holographic dark energy. We hope we will return to this
issue in future projects.

We also discussed the coincidence problem. It is shown that the
interacting holographic dark energy approach only solves the
coincidence problem partially.  The original solution to the coincidence
problem proposed in \cite{mli04} stands a better resolution.

\section*{Acknowledgments}
We thank Qing-Guo Huang, Tower Wang for discussions. This work was
supported by grants from NSFC, a grant from Chinese Academy of
Sciences and a grant from USTC.

\end{document}